\documentclass[twocolumn,twocolappendix]{aastex631}

\usepackage{graphicx,enumitem}
\usepackage{xcolor,graphicx,xspace,color,longtable,}
\usepackage{amssymb,amsmath,shadow,bezier,curves,rotating}
\usepackage{graphicx,chngcntr}
\graphicspath{ {./images/}}
\usepackage{color,soul}
\usepackage{booktabs}
\usepackage[flushleft]{threeparttable}
\usepackage[autostyle=true]{csquotes}
\usepackage[english]{babel}
\usepackage{hyperref}
\usepackage{tabularx} 
\setlist[enumerate]{leftmargin=0pt,labelindent=0pt}

\newcommand {\ks}   {$\mathrm{km}~\mathrm{s}^{-1}$}
\newcommand {\hm}   {$h^{-1} \  \mathrm{M}_{\odot}$}

\defcitealias{Biviano16}{Biv16}
\defcitealias{Biviano21}{Biv21}

\DeclareSymbolFont{matha}{OML}{txmi}{m}{it}
\DeclareMathSymbol{\varv}{\mathord}{matha}{118}

\begin{document}

\title[Detection and Dynamics of Galaxy Clusters in VIPERS Survey]
{Galaxy Cluster Detection and Dynamical Analysis in the VIPERS High-Redshift Spectroscopic Survey}

\author[0000-0003-3595-7147]{Mohamed H. Abdullah}
\affiliation{Department of Physics, University of California Merced, 5200 North Lake Road, Merced, CA 95343, USA}
\affiliation{Department of Astronomy, National Research Institute of Astronomy and Geophysics, Cairo, 11421, Egypt}

\author[0009-0004-4674-6446]{Omnia I. Adly}
\affiliation{Department of Physics, University of California Merced, 5200 North Lake Road, Merced, CA 95343, USA}

\author[0000-0002-6572-7089]{Gillian Wilson}
\affiliation{Department of Physics, University of California Merced, 5200 North Lake Road, Merced, CA 95343, USA}

\author[0009-0008-5349-5410]{Magdy Y. Amin}
\affiliation{Department of Astronomy, Space Science, and Meteorology, Faculty of Science, Cairo University, Giza 11326, Egypt}

\author[0000-0001-9070-4914]{A. Ahmed}
\affiliation{Department of Astronomy, Space Science, and Meteorology, Faculty of Science, Cairo University, Giza 11326, Egypt}

\begin{abstract}
We present a dynamical analysis of galaxy clusters identified in the VIPERS spectroscopic survey within the redshift range \(0.5 \leq z \leq 1.2\). Cluster candidates were first detected as overdense regions in redshift space through the Finger-of-God (FoG) effect, and cluster membership was assigned using the GalWeight technique within the FoG–GalWeight methodology developed by our team. For each cluster, we derived the virial radius (\(R_{200}\)), velocity dispersion (\(\sigma_{200}\)), and virial mass (\(M_{200}\)) using the virial mass estimator. We identified ten VIPERS clusters spanning a mass range of \(0.59\times10^{14} \leq M_{200}/(h^{-1}M_\odot) \leq 4.32\times10^{14}\) and velocity dispersions of \(360 \lesssim \sigma_{200} \lesssim 900\)~km~s\(^{-1}\). 
We cross-matched the VIPERS clusters with published catalogs and found at least one matching system for each cluster, offering external validation for our detections.
We investigated the velocity dispersion–mass relation $\sigma$MR for these systems and obtained $\log{\sigma_{200}} = (2.73 \pm 0.06) + (0.36 \pm 0.18)\,\log{M_{200}}$, with an intrinsic scatter of \(\sigma_\mathrm{int} = 0.04 \pm 0.07\). The derived relation is consistent with theoretical predictions from \(N\)-body and hydrodynamical simulations, confirming the reliability of the FoG–GalWeight methodology and the robustness of the virial mass estimator. Our findings demonstrate that the velocity dispersion can serve as a reliable and direct proxy for cluster mass, even at high redshift, without requiring additional dynamical mass modeling.
\end{abstract}

\section{Introduction} \label{sec:intro}

Galaxy clusters represent the most massive gravitationally bound systems in the Universe, marking the densest nodes of the cosmic web where dark matter, gas, and galaxies coevolve \citep{Carlberg96,Allen11,Kravtsov12}. As powerful cosmological probes, they trace the growth of large-scale structure and provide stringent constraints on cosmological parameters such as $\Omega_\mathrm{m}$ and $\sigma_8$ \citep{Reiprich02,Kravtsov12,Abdullah24}. At the same time, clusters serve as astrophysical laboratories for studying galaxy evolution in dense environments, where processes such as ram-pressure stripping, galaxy--galaxy interactions, and quenching are particularly efficient \citep{Tyler13,Ebeling14,Boselli14,Tagliaferro21}.

The study of high-redshift galaxy clusters is especially valuable for understanding when and how these massive systems formed. Clusters observed at $0.5 \lesssim z \lesssim 1.2$ are caught in an era of active assembly, when the balance between accretion and relaxation is still evolving \citep{Overzier16,McConachie22}. Their dynamical properties, such as velocity dispersion, substructure, and total mass, encode information about both cosmological growth and internal relaxation processes. However, identifying and characterizing these systems at high redshift remains observationally challenging due to the limited sampling of spectroscopic data and the projection effects inherent to photometric catalogs.  

Large spectroscopic surveys such as the VIMOS Public Extragalactic Redshift Survey (VIPERS) \citep{Guzzo14,Scodeggio18} now provide an unprecedented opportunity to overcome these challenges. With accurate redshifts for over 90,000 galaxies in the range $0.5 < z < 1.2$, VIPERS offers the statistical depth necessary to detect and study galaxy clusters in the distant Universe through their dynamical signatures. Yet, a major difficulty in exploiting such datasets lies in identifying clusters and assigning reliable memberships directly from redshift space, without relying on pre-existing cluster catalogs or photometric overdensity searches.

In this work, we apply the FoG–GalWeight methodology \citep{Abdullah18,Abdullah20a}, a hybrid approach developed to identify and characterize galaxy clusters directly from spectroscopic data. The first component of the method exploits the Finger-of-God (FoG) effect, a characteristic elongation of galaxy distributions along the line of sight caused by peculiar velocities within gravitationally bound systems. This distortion in redshift space serves as a robust indicator of overdense regions associated with galaxy clusters, enabling their detection even in surveys with limited angular coverage such as VIPERS. By mapping these redshift-space overdensities, the FoG component effectively isolates candidate cluster regions for subsequent dynamical analysis.

The second component, the GalWeight technique \citep{Abdullah18}, refines cluster membership within these FoG-detected overdensities. GalWeight employs a dynamical weighting scheme in projected phase space ($R_p$, $v_\mathrm{pec}$) to statistically distinguish true members from foreground and background interlopers. Unlike traditional membership algorithms such as the 3$\sigma$-clipping \citep{Yahil77}, gapper \citep{Beers90,Zabludoff90}, and shifting-gapper \citep{Fadda96} methods, which rely on fixed selection thresholds or iterative cuts, GalWeight adaptively maximizes completeness while minimizing contamination, achieving over 98\% accuracy in simulation tests. This dual FoG–GalWeight methodology thus enables both the reliable detection and dynamical characterization of galaxy clusters in large spectroscopic surveys such as VIPERS.

Unlike many photometric or statistical cluster-finding techniques that rely on color–magnitude relations or predefined richness thresholds, the FoG–GalWeight approach uses purely spectroscopic information to identify gravitationally bound systems. This makes it particularly valuable for high-redshift surveys, where photometric uncertainties and projection effects can bias cluster detection. By relying on dynamical and redshift-space information alone, the method provides a physically motivated route for constructing spectroscopic cluster catalogs with minimal contamination.

The FoG–GalWeight methodology has already been validated in both simulated and observational contexts, including its application to the SDSS spectroscopic data that produced the GalWCat19 catalog \citep{Abdullah18,Abdullah20a}. These studies demonstrated the method’s efficiency in identifying clusters, determining accurate dynamical masses, and constraining cosmological parameters \citep{Abdullah20b,Abdullah23,Abdullah24}. In this work, we extend the framework to the VIPERS spectroscopic survey to probe an earlier epoch of cluster formation and evaluate its performance in a more challenging high-redshift regime. Although VIPERS was not originally designed for cluster detection, its high redshift precision and dense sampling enable the recovery of a modest yet robust sample of dynamically confirmed systems. Our goal is to assess whether gravitationally bound structures can be reliably identified and characterized directly from the spectroscopic data using this fully dynamical approach. Through this analysis, we search for redshift-space overdensities, determine their memberships with the GalWeight technique, and derive their key dynamical properties. This study therefore represents the first high-redshift application of the FoG–GalWeight methodology, bridging the gap between current spectroscopic cluster catalogs (e.g., GalWCat19) and those anticipated from next-generation surveys such as DESI and \textit{Euclid}.

Once cluster membership is established, we determine their dynamical properties using the virial mass estimator \citep{Rines03,Abdullah20a}, applying corrections for the surface pressure term \citep{The86,Carlberg97}. This approach has been validated against hydrodynamical and N-body simulations and shown to perform comparably to, or better than, other complex techniques  \citep{Old15,Abdullah20a}. By measuring the line-of-sight velocity dispersion ($\sigma_\mathrm{los}$) and total dynamical mass ($M_{200}$), we also explore the velocity dispersion--mass relation ($\sigma$MR), an important empirical scaling that connects observable kinematics to total mass and provides an efficient route for estimating cluster masses in large datasets.

The paper is organized as follows. In Section~\ref{sec:data}, we describe the VIPERS spectroscopic dataset, the methodology for detecting clusters and their membership using the FoG–GalWeight methodology, and the procedures for calculating their dynamical properties. Section~\ref{sec:results} presents the identified cluster sample and their derived physical parameters. In Section~\ref{sec:sigmamass}, we examine $\sigma$MR for the VIPERS clusters and compare our results with theoretical predictions from cosmological simulations. Finally, our conclusions are summarized in Section~\ref{sec:conc}. Throughout this paper, we adopt a flat \(\Lambda\)CDM cosmology consistent with the \citet{Planck15} results, assuming \(\Omega_{\mathrm{M}} = 0.3089\), \(\Omega_{\Lambda} = 0.6911\), and \(h = 0.6774\). All logarithms are base-10 (\(\log_{10}\)) unless otherwise specified.

\section{Data and Methodology} \label{sec:data}
In this section, we describe the dataset and the methodology used to identify and analyze galaxy clusters in the VIPERS survey. We first outline the spectroscopic data and its key characteristics, then detail the cluster-finding procedure based on detecting overdensities in redshift space through the Finger-of-God (FoG) effect and assigning cluster membership using the GalWeight technique. Finally, we present the dynamical analysis method employed to derive the physical properties of each identified cluster, including the virial radius ($R_{200}$), velocity dispersion ($\sigma_v$), and total mass ($M_{200}$).

\subsection{VIPERS Sample}\label{sec:VIPERS}
In this study, we utilize data from VIPERS\footnote{\url{http://vipers.inaf.it/rel-pdr2.html}}, a large spectroscopic survey designed to investigate galaxy clustering, dynamics, and evolution in the redshift range \(0.5 \leq z \leq 1.2\) \citep{Guzzo14,Scodeggio18}. The survey covers two primary fields, W1 and W4, selected from the five-band catalogs of the Canada–France–Hawaii Telescope Legacy Survey (CFHTLS)–Wide \citep{Guzzo14,Garilli14}. Together, these fields span approximately 24~deg$^2$, corresponding to a comoving volume of $\sim 5\times10^7\,h^{-3}\,\mathrm{Mpc}^3$ with a median redshift of $z \sim 0.8$. The dataset contains nearly 90,000 galaxies, providing a statistically robust foundation for studies of large-scale structure and galaxy environments \citep{Scodeggio18}.

VIPERS applied a magnitude limit of $i_{\rm AB} < 22.5$, achieving a spectroscopic completeness of approximately 88\% for targets with secure redshift flags. Target selection and slit placement were optimized using the SPOC (Slit Positioning and Optimization Code) algorithm implemented within the VIMOS multi-object spectrograph \citep{Garilli14}. Spectroscopic redshifts were measured with a typical uncertainty of $\sigma_z \approx 0.00054(1+z)$ for high-confidence sources \citep{Scodeggio18,Pezzotta17}, with quality flags $\geq2$ corresponding to a confidence level above 96\% \citep{Scodeggio18}.

For this analysis, we select galaxies within $0.5 \leq z \leq 1.2$ with reliable spectroscopic redshifts (flag $\geq 2$) and classified as primary targets. For each galaxy, we retrieve its right ascension ($\alpha$), declination ($\delta$), and spectroscopic redshift ($z$). The final sample used in this work is based on the complete Second Public Data Release (PDR-2; \citealt{Scodeggio18}).

\subsection{Cluster-Finding Technique} \label{sec:finding}
Galaxy clusters are identified as overdense regions in redshift space whose galaxy densities exceed the cosmic background by two to three orders of magnitude. A characteristic feature of clusters in redshift space is the distortion of galaxy velocities along the line of sight near the cluster center (within $\sim0.5$~Mpc), known as the Finger-of-God (FoG) effect \citep{Jackson72,Kaiser87,Abdullah13}. This effect arises from the random motions of galaxies within the gravitational potential well and is clearly visible in the projected phase-space diagram of line-of-sight velocity ($v_\mathrm{pec}$) versus projected radius ($R_p$), where
\begin{equation} \label{eq:01}
v_\mathrm{pec} = \frac{v_{\mathrm{obs}} - v_c}{1 + z_c},
\end{equation}
where $v_{\mathrm{obs}}$ is the observed spectroscopic velocity of the galaxy, and $z_c$ and $v_c$ represent the cluster redshift and systemic velocity, respectively. The $(1 + z_c)$ term corrects for the cosmological Hubble expansion \citep{Danese80}. 

Cluster detection and membership assignment are performed using the FoG–GalWeight methodology \citep{Abdullah18,Abdullah20a}, which combines FoG-based overdensity identification with the GalWeight membership technique. The procedure proceeds as follows:
\begin{enumerate}[leftmargin=*]
    \item Local density estimation: For each galaxy $i$, we calculate the local number density, $\rho_\mathrm{cy}$, within a cylindrical volume centered on that galaxy, with a radius of $R_\mathrm{cy} = 1.0~h^{-1}$~Mpc (approximately the typical width of the FoG) and a height of $3000$~km~s$^{-1}$ (the typical FoG length). This radius corresponds to an angular scale $\sin(\theta_\mathrm{cy}) = R_\mathrm{cy}/D_\mathrm{c,g}$, where $D_\mathrm{c,g}$ is the comoving distance to the galaxy, given by
    \begin{equation}
        D_\mathrm{c,g} = \frac{c}{H_0} \int_{0}^{z} \frac{1}{\sqrt{\Omega_\mathrm{m}(1+z')^3 + \Omega_\mathrm{k}(1+z')^2 + \Omega_\Lambda}} \, dz' 
    \end{equation}
    \item Cluster candidate identification: All galaxies are ranked in descending order of local density, requiring at least seven galaxies within a projected radius of $R_\mathrm{cy} = 1.0~h^{-1}$~Mpc and a velocity range of $\pm1500$~km~s$^{-1}$ from the center. This ensures the detection of statistically significant overdensities suitable for cluster identification.
    \item Cluster center determination: For each high-density region, a binary tree algorithm \citep{Serra11} and the two-dimensional Adaptive Kernel Method (2DAKM; e.g., \citealp{Pisani96}) are applied to locate the cluster center coordinates ($\alpha_c$, $\delta_c$), while the one-dimensional Adaptive Kernel Method (1DAKM) is used to determine the cluster redshift ($z_c$). This provides a robust definition of the cluster center and allows construction of the projected phase-space diagram ($R_\mathrm{p}$, $v_\mathrm{pec}$).
    \item Membership assignment using GalWeight: The GalWeight technique \citep{Abdullah18} is then applied to galaxies within $R_{\mathrm{p,max}} = 8~h^{-1}$~Mpc and a maximum velocity offset of $v_{z,\mathrm{max}} = 3000$~km~s$^{-1}$. GalWeight identifies the cluster boundary in redshift space by assigning dynamical weights to galaxies based on their position in projected phase space. The method effectively separates true cluster members from foreground and background interlopers, minimizing contamination while maintaining high completeness. GalWeight has been validated using both simulated and observed datasets and has been shown to recover cluster membership with high accuracy \citep{Abdullah18,Abdullah20a}.
\end{enumerate}

This FoG–GalWeight methodology provides a robust, fully spectroscopic approach to detecting and defining galaxy clusters in the VIPERS dataset. Unlike traditional two-dimensional or photometric cluster-finding methods that rely solely on projected galaxy densities, this method exploits the full three-dimensional redshift-space information, allowing the identification of both rich and poor clusters with high confidence \citep{Abdullah20a}.

\subsection{Dynamical Analysis Method} \label{sec:dynamics}
\label{sec:dynamics}
The total mass of a galaxy cluster can be estimated using the virial mass estimator (e.g., \citealp{Limber60,Binney87,Rines03}) or by assuming an NFW density profile \citep{NFW96,NFW97}. The virial mass estimator is given by
\begin{equation} \label{eq:vir1}
M_{VT} = \frac{3\pi N \sum_{i} v_{\mathrm{pec},i}^2}{2G \sum_{i\neq j} \frac{1}{R_{ij}}},
\end{equation}
where $v_{\mathrm{pec},i}$ is the galaxy line-of-sight peculiar velocity in the cluster rest frame (see Equation~\ref{eq:01}), and $R_{ij}$ is the projected separation between galaxies $i$ and $j$. Here, $v_{\mathrm{obs}}$ is the observed velocity of the galaxy, and $z_c$ and $v_c$ represent the cluster redshift and systemic velocity, respectively. Equation~(\ref{eq:vir1}) holds only within the virialized region, which extends approximately to the virial radius $R_{200}$.

Since galaxy clusters extend beyond the virial radius, Equation~(\ref{eq:vir1}) tends to overestimate the total mass due to the contribution of external pressure from material outside the virialized region \citep{The86,Carlberg97,Girardi98a}. The corrected virial mass is therefore expressed as
\begin{equation} \label{eq:vir2}
M_{VTC} = M_{VT}\,[1 - S(r)],
\end{equation}
where $S(r)$ represents the surface pressure correction term.  

For a cluster following an NFW density profile and assuming isotropic orbits (i.e., $\sigma_v = \sigma_\theta$, or equivalently, the velocity anisotropy parameter $\beta = 1 - \sigma_\theta^2/\sigma_r^2 = 0$), the correction term can be written as \citep{Koranyi00,Abdullah11}
\begin{equation} \label{eq:vir_25}
S(r) = 
\left( \frac{x}{1+x} \right)^2
\left[ \ln(1+x) - \frac{x}{1+x} \right]^{-1}
\left[ \frac{\sigma_v(r)}{\sigma(<r)} \right]^2,
\end{equation}
where $x = r/r_s$, $r_s$ is the NFW scale radius, $\sigma(<r)$ is the integrated three-dimensional velocity dispersion within radius $r$, and $\sigma_v(r)$ is the projected velocity dispersion at $r$.

The NFW mass density profile is defined as
\begin{equation} \label{eq:NFW1}
\rho(r) = \frac{\rho_s}{x(1+x)^2},
\end{equation}
and the corresponding enclosed mass profile is
\begin{equation} \label{eq:NFW11}
M(<r) = \frac{M_s}{\ln(2) - \tfrac{1}{2}}
\left[ \ln(1+x) - \frac{x}{1+x} \right],
\end{equation}
where $M_s = 4\pi\rho_s r_s^3 [\ln(2) - \tfrac{1}{2}]$ is the mass within $r_s$, and $\rho_s = \delta_s \rho_c$ is the characteristic density. The overdensity parameter $\delta_s$ is given by
\[
\delta_s = \frac{\Delta_{200}}{3} c^3 
\left[\ln(1+c) - \frac{c}{1+c}\right]^{-1},
\]
where $c = R_{200}/r_s$ is the concentration parameter 
(e.g., \citealp{NFW97,Rines03,Mamon13}).  

\begin{table*}
\caption{Coordinates, redshifts, and dynamical properties of the ten VIPERS clusters.}
\label{tab:pos}
\begin{tabular}{ccccccccc}
\hline
Cluster ID   &Field&$\alpha_c$&$\delta_c$ &$z_c$ &$R_{200}$ &$N_{200}$ &$\sigma_{200}$  &$M_{200}$\\
&&(deg)  &(deg)& &$(h^{-1}$~Mpc)   &           &(km s$^{-1}$)     &($10^{14}~h^{-1}~\mathrm{M}_{\odot}$)\\
\hline
VIPERS01 & W4 &333.804 &  1.044 & 0.700 & 0.945 &  8 & $902_{-301}^{+353}$ & $4.324 \pm 0.963$ \\
VIPERS02 & W1&  30.766 & -5.017 & 0.509 & 0.964 & 21 & $710_{-156}^{+125}$ & $3.651 \pm 0.621$ \\
VIPERS03 & W1&  37.126 & -4.660 & 0.637 & 0.708 &  9 & $679_{-251}^{+317}$ & $1.687 \pm 0.368$ \\
VIPERS04 & W1&  35.811 & -4.959 & 0.826 & 0.641 &  8 & $500_{-147}^{+114}$ & $1.600 \pm 0.356$ \\
VIPERS05 & W1&  34.696 & -5.139 & 0.552 & 0.674 &  7 & $503_{-187}^{+239}$ & $1.313 \pm 0.308$ \\
VIPERS06 & W4& 334.192 &  1.781 & 0.567 & 0.661 &  8 & $474_{-191}^{+201}$ & $1.263 \pm 0.287$ \\
VIPERS07 & W1&  37.734 & -5.376 & 0.622 & 0.604 &  8 & $466_{-145}^{+106}$ & $1.037 \pm 0.237$ \\
VIPERS08& W1&  32.470 & -5.697 & 0.538 & 0.626 &  9 & $478_{-153}^{+288}$ & $1.033 \pm 0.223$ \\
VIPERS09 & W1&  31.624 & -5.431 & 0.606 & 0.594 &  7 & $490_{-254}^{+153}$ & $0.951 \pm 0.219$ \\
VIPERS10& W4 & 330.601 &  1.503 & 0.582 & 0.500 &  7 & $363_{-126}^{+149}$ & $0.591 \pm 0.138$ \\
\hline
\end{tabular}
\end{table*}

The projected number of galaxies within a cylinder of radius $R$ can be obtained by integrating Equation~(\ref{eq:NFW1}) along the line of sight (e.g., \citealp{Bartelmann96,Zenteno16}):
\begin{equation} \label{eq:NFW2}
N(<R) = \frac{N_s}{\ln(2) - \tfrac{1}{2}}\, g(x),
\end{equation}
where $N_s$ is the number of galaxies within $r_s$ (analogous to $M_s$), and $g(x)$ is given by (e.g., \citealp{Golse02,Mamon10})
\begin{equation}
g(x) = 
\begin{cases}
\ln(x/2) + \frac{\cosh^{-1}(1/x)}{\sqrt{1 - x^2}}, & x < 1, \\[8pt]
1 - \ln(2), & x = 1, \\[8pt]
\ln(x/2) + \frac{\cos^{-1}(1/x)}{\sqrt{x^2 - 1}}, & x > 1. 
\end{cases}
\end{equation}

The virial theorem is strictly valid only within the virialized region. In our analysis, we first fit the NFW scale radius $r_s$ to compute the surface pressure correction term $S(r)$. We then determine the virial radius $R_{200}$ by requiring that the mean enclosed density equals $200\,\rho_c$, where $\rho_c$ is the critical density of the Universe at the cluster redshift. The cluster virial mass is then defined as the corrected virial mass $M_{VTC}$ enclosed within this radius, which we also refer to as $M_{200}$ throughout this paper.

Applying this methodology to the VIPERS spectroscopic dataset, we identified ten galaxy clusters spanning the redshift range $0.5 < z < 0.85$. The results confirm that, despite the survey’s original design not being optimized for cluster detection, the FoG–GalWeight methodology is capable of reliably identifying and characterizing dynamically bound systems in high-redshift spectroscopic data. The derived cluster properties and individual parameters are presented in Section~\ref{sec:results}.
\begin{figure*}\hspace{-0cm}  \includegraphics[width=0.9\linewidth]{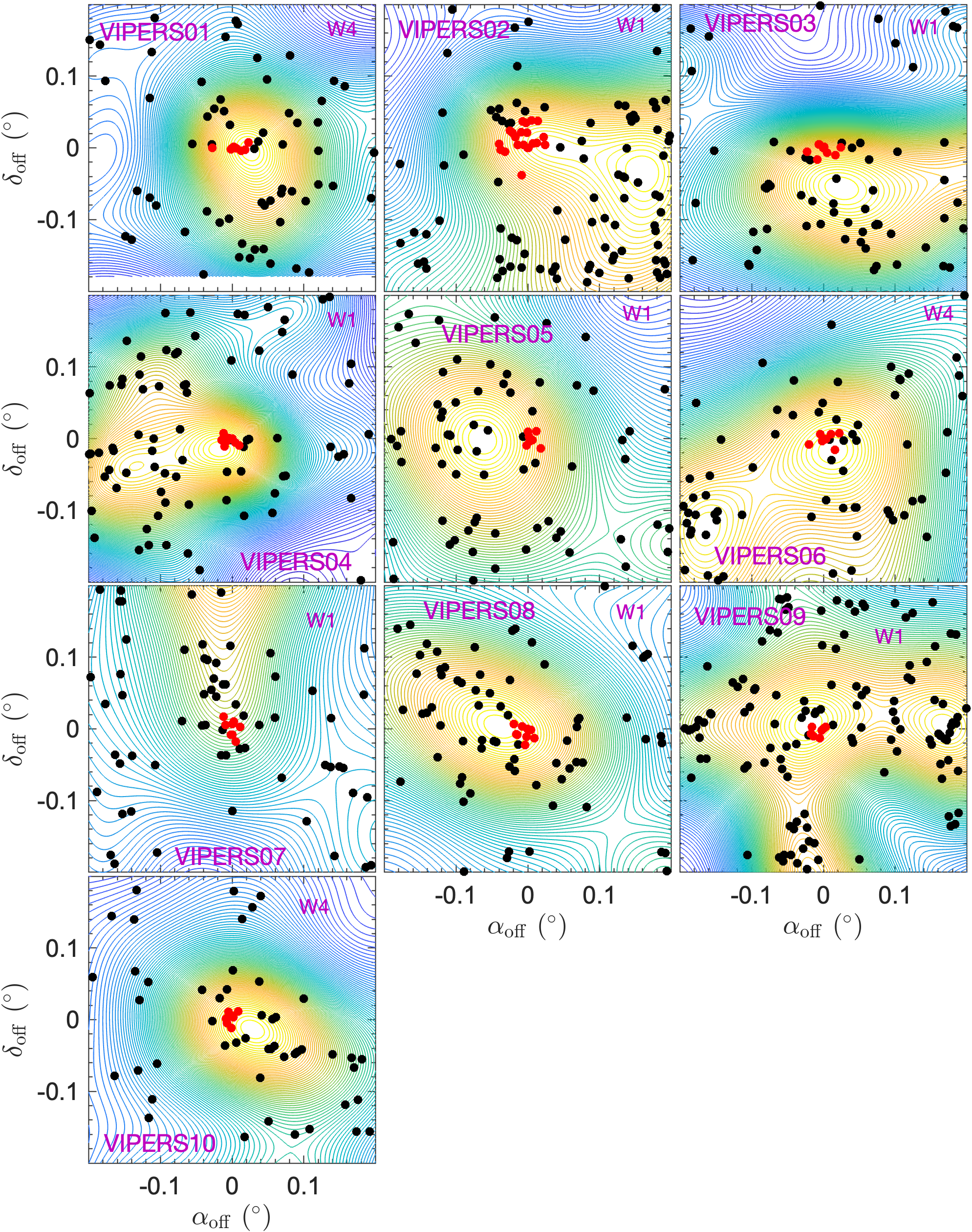} \vspace{-.0cm}
\centering
    \caption{Projection maps of the VIPERS clusters. Black points show the offsets ($\alpha_\mathrm{off}$, $\delta_\mathrm{off}$) from the cluster center for all spectroscopically confirmed galaxies (members and non-members) within a velocity window of $\Delta v = \pm 3500$~km~s$^{-1}$ from each cluster redshift (see Section~\ref{sec:finding}). Red points mark the spectroscopically confirmed member galaxies that satisfy these criteria and also lie within $R_{200}$ (see also Figure~\ref{fig:PS}, which shows their distribution in projected phase-space diagram). The contours were computed using the two-dimensional adaptive kernel method (2DAKM; e.g., \citealp{Pisani96}) and trace the surface density of all (red and black) spectroscopically confirmed galaxies.}
   \label{fig:Coors}
\end{figure*}

\section{VIPERS Cluster Sample and Their Properties} 
\label{sec:results}

In this section, we present the VIPERS cluster sample identified using the FoG--GalWeight methodology and summarize their derived physical properties. The procedures used to determine the cluster centers, assign spectroscopic membership, and compute the dynamical quantities are described in Sections~\ref{sec:finding} and~\ref{sec:dynamics}.

\subsection{VIPERS cluster sample}
\label{sec:clustersample}

The FoG--GalWeight methodology identifies ten galaxy clusters within the VIPERS W1 and W4 fields. Table~\ref{tab:pos} lists the coordinates of the cluster centers [($\alpha_c$, $\delta_c$)] and their redshifts ($z_c$), along with the derived dynamical quantities: $R_{200}$, $N_{200}$, $\sigma_{200}$, and the corresponding virial mass $M_{200}$. The uncertainty in $\sigma_{200}$ was estimated using a bootstrap resampling technique. For each cluster, the velocity dispersion was recalculated from 1,000 random resamplings of its member galaxies, and the $1\sigma$ confidence interval of the resulting distribution was adopted as the uncertainty. Given the relatively small number of spectroscopic members in some clusters, the velocity dispersion itself was computed using the Gapper estimator \citep{Beers90}, which provides a robust measure for samples with limited galaxy counts. The uncertainty in $M_{200}$ was derived by propagating the corresponding error in $\sigma_{200}$ through the virial mass estimator and by including the limiting fractional uncertainty of the virial mass estimator, given by (\citealp{Bahcall81}
\begin{equation}
\frac{\Delta M}{M} = \pi^{-1} (2 \ln N)^{1/2} N^{-1/2}
\end{equation}

The identified VIPERS clusters span the redshift range $0.5 < z < 0.85$, with virial radii between $0.5$ and $1.0~h^{-1}$~Mpc and velocity dispersions of $360$–$900$~km~s$^{-1}$. The corresponding virial masses lie in the range $(0.6$–$4.3)\times10^{14}~h^{-1}\,M_\odot$. These values indicate that the detected systems represent a mixture of moderately rich groups and low- to intermediate-mass clusters, characteristic of the VIPERS redshift range.

\begin{figure*}\hspace{-0cm}  \includegraphics[width=0.9\linewidth]{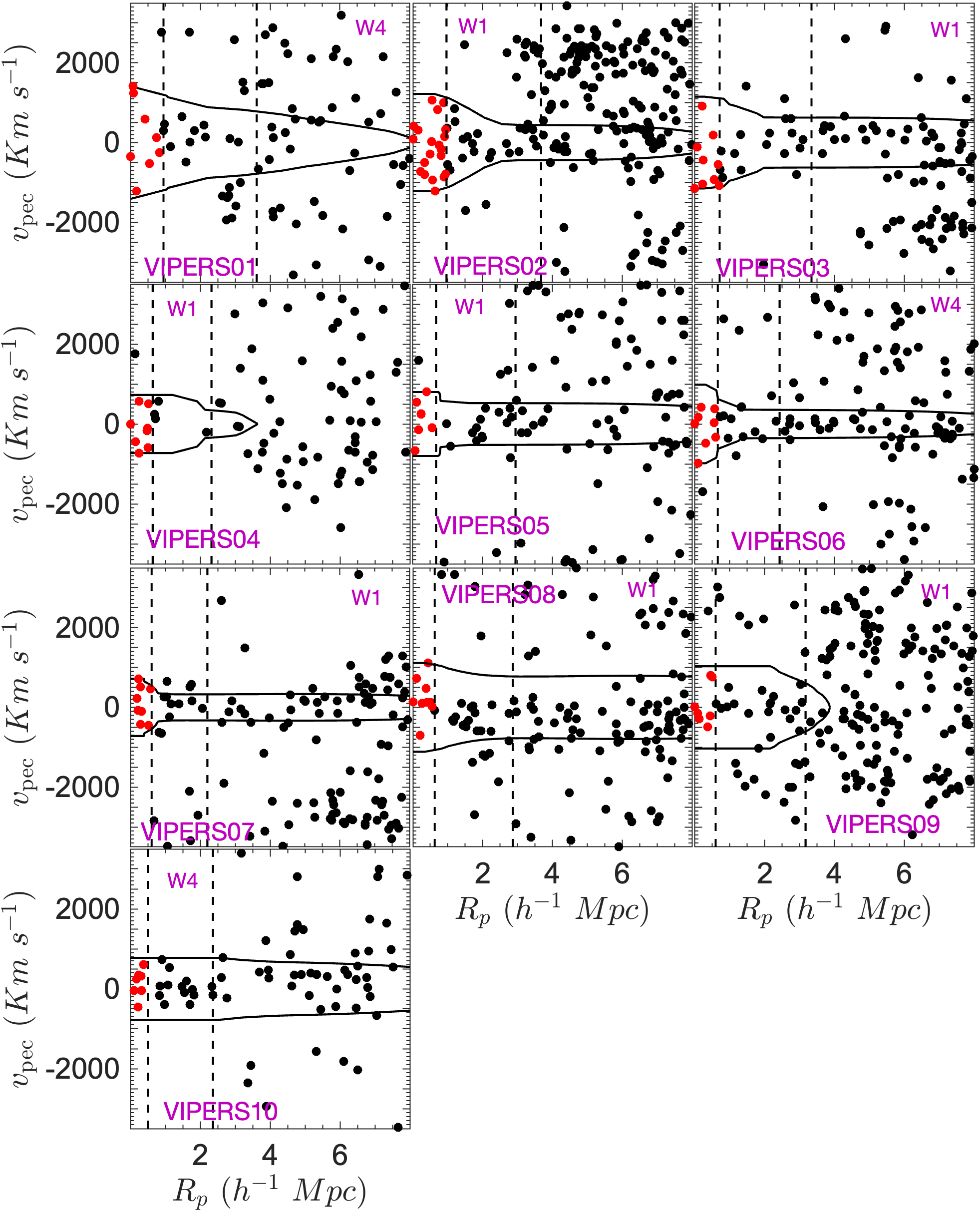} \vspace{-0.1cm}
\centering
     \caption{Projected phase-space diagrams where $\mathrm{v_{pec}}$ and $R_p$ are the peculiar velocity and the projected distance, respectively. Application of the GalWeight technique results in the solid black line (cluster boundary) within which are enclosed spectroscopically-confirmed galaxies which are identified as cluster members. Black points show all spectroscopically-confirmed galaxies within the field of view (shown also in Figure~\ref{fig:Coors}) and red points indicate cluster members identified by the GalWeight technique (black contour) but which also fall within $R_{200}$. The two vertical dashed lines correspond to the virial radius $R_{200}$ and the turnaround radius $R_\mathrm{ta}$.}
   \label{fig:PS}
\end{figure*}

Figure~\ref{fig:Coors} shows the projected distribution of galaxies around each VIPERS cluster. Offsets in right ascension and declination, $(\alpha_\mathrm{off}, \delta_\mathrm{off})$, are measured relative to the cluster center. Black points represent all spectroscopically confirmed galaxies (members and non-members) within a line-of-sight velocity window of $\Delta v = \pm 3500$~km~s$^{-1}$ from the cluster redshift. Red points indicate the spectroscopically confirmed member galaxies that satisfy the GalWeight membership criteria and lie within the virial radius $R_{200}$. Contours were generated using the two-dimensional adaptive kernel method (2DAKM; e.g., \citealp{Pisani96}) and trace the surface density of all spectroscopically confirmed galaxies in each field. The highest contour levels correspond to the cluster cores, where galaxy density peaks. For comparison, Figure~\ref{fig:PS} shows the same galaxies in projected phase space ($v_\mathrm{pec}$ vs.\ $R_p$).

\begin{table*}
\centering
\begin{threeparttable}
\caption{Cross-matching of the VIPERS clusters with previously published cluster catalogs. Coordinates and redshifts of the VIPERS clusters are listed in Table~1. Here we list the matched clusters identified in the literature.}
\label{tab:crossmatch}

\footnotesize        

\begin{tabular}{@{}llllllll@{}}   
\hline
\shortstack{VIPERS\\Cluster} &
\shortstack{Matched \\Cluster Name} &
\shortstack{$\alpha$\\(deg)} &
\shortstack{$\delta$\\(deg)} &
\shortstack{$z$} &
\shortstack{Sep.\\(arcmin)} &
\shortstack{$|{\Delta z}|$} &
\shortstack{Reference} \\
\hline
VIPERS01 & HSCS J221514+010241       & 333.809 &  1.045 & 0.699 & 0.341 & 0.000 & \citet{Oguri18} \\

VIPERS02 & CFHTLS:[SMD2018a] W1-1066 &  30.750 & -5.004 & 0.500 & 1.204 & 0.009 & \citet{Sarron18} \\
         & WHL J020302.2-045938 &  30.759 & -4.994 & 0.509 & 1.448 & 0.001 & \citet{Wen15} \\       

VIPERS03 & CFHT-W CL J022830.0-044413 &	37.131	& -4.734	& 0.640 & 4.463 & 0.003 & \citet{Wen11} \\

VIPERS04 & CFHT-W CL J022315.2-045737  & 35.797 & -4.961 & 0.845  & 0.867 & 0.018 &\citet{Wen11}\\

VIPERS05 & CFHT-W CL J021847.0-050857  & 34.701 & -5.140 & 0.580  & 0.330 & 0.028 &\citet{Wen11}\\

VIPERS06 & HSCS J221647+014648  & 334.196  & 1.780 & 0.568 & 0.225 & 0.001 & \citet{Oguri18}\\ 
         & WHL J221646.9+014648 & 334.196  & 1.780 & 0.567 & 0.225 & 0.000 & \citet{Wen15}\\
         & WHL J221646.9+014648 & 334.184  & 1.727 & 0.567 & 3.266 & 0.013 & \citet{Durret11}\\
         
VIPERS07 & CFHTLS:[DAC2011] W1-2861      &  37.721 & -5.406 & 0.610 & 2.011 & 0.012 &\citet{Durret11}\\

VIPERS08 & WHL J020949.8-054210          & 32.458 & -5.703 & 0.549 & 0.817 & 0.011 &\citet{Wen21} \\
         & CFHTLS:[SMD2018a] W1-1277 &  32.448 & -5.722 & 0.550 & 2.000 & 0.012 &\citet{Sarron18}\\

VIPERS09 & WHL J020627.7-052600      & 31.615 & -5.434 & 0.608 & 0.553 & 0.003 &\citet{Wen15}\\
 		 & CFHTLS:[SMD2018a] W1-1441& 31.611 & -5.436 & 0.590 & 0.879 & 0.016 &\citet{Sarron18}\\
 		 & CFHTLS:[DAC2011] W1-0580 & 31.596 & -5.434 & 0.590 & 1.702 & 0.016 &\citet{Durret11}\\

VIPERS10 & CFHTLS:[DAC2011] W4-0086  & 330.589&  1.544 & 0.570 & 0.743 &  0.012 & \citet{Durret11} \\
\hline
\end{tabular}
\end{threeparttable}
\end{table*}

Figure~\ref{fig:PS} shows the projected phase-space diagrams ($v_\mathrm{pec}$ versus $R_p$) for the spectroscopically confirmed galaxies shown in Figure~\ref{fig:Coors}. Black points represent all galaxies within $\Delta v = \pm3500$~km~s$^{-1}$ of each cluster redshift, while red points indicate spectroscopically confirmed member galaxies that satisfy the GalWeight membership criteria and lie within $R_{200}$. The solid black curves mark the optimal GalWeight caustics, which delineate the cluster boundaries in projected phase space. The vertical dashed lines correspond to the virial radius $R_{200}$ and the turnaround radius $R_\mathrm{ta}$. The turnaround radius $r_\mathrm{ta}$ is the radius at which a galaxy’s peculiar velocity ($v_\mathrm{pec}$) is canceled by the global Hubble expansion. In other words, it is the radius where the infall velocity vanishes ($v_\mathrm{inf} = v_\mathrm{pec} - H r = 0$), which can be calculated as the radius at which $\rho = 5.55\,\rho_\mathrm{c}$ (e.g., \citealp{Nagamine03,Busha05,Dunner06}).

In redshift space, galaxy clusters exhibit a characteristic “trumpet-shaped” pattern: the velocity dispersion is highest near the cluster center and decreases with increasing projected radius (\citealp{Regos89,Praton94,Diaferio99,Abdullah13}). This behavior reflects the stronger gravitational binding in the central regions, where galaxies move faster within the cluster potential well, and the progressively lower peculiar velocities in the outer regions, where galaxies are less bound. The resulting narrowing of the velocity distribution at large radii, clearly visible in these diagrams, is consistent with both simulations and observations (e.g., \citealp{Abdullah18,Paez22}). The figure demonstrates that the GalWeight technique accurately traces these caustic profiles, efficiently separating cluster members from foreground and background galaxies, even for clusters with a small number of galaxies. Tests with simulations show that GalWeight correctly identifies more than 98\% of true members and outperforms several widely used membership methods, including the shifting gapper, den Hartog, caustic, and Spherical Infall techniques \citep{Abdullah18}.

The performance of the virial mass estimator method in recovering cluster mass was assessed using two mock catalogs, HOD2 and SAM2, as referenced in \citet{Abdullah20a}. They employed various statistics to examine the performance of the mass reconstruction methods. These statistics include the root-mean-square ($\mathrm{rms}$) difference between the recovered and true log mass, the scatter in the recovered mass ($\sigma_{\mathrm{M}_\mathrm{rec}}$), the scatter about the true mass ($\sigma_{\mathrm{M}_\mathrm{true}}$), and the bias at the pivot mass (taken as the median log mass of the input cluster sample) between the recovered and the true mass (see \citealp{Old15} and the references therein). Specifically, $\mathrm{rms}$, $\sigma_{\mathrm{M}_\mathrm{rec}}$, $\sigma_{\mathrm{M}_\mathrm{{true}}}$, and the bias values for the virial mass estimator returned \(0.24\), \(0.23\), \(0.23\), and \(0.06\) for HOD2, and \(0.32\), \(0.21\), \(0.23\), and \(0.24\) for SAM2, respectively \citep{Abdullah20a}. 

\subsection{Cross-matching with previously identified clusters}
\label{sec:crossmatch}

To determine whether the VIPERS clusters correspond to systems previously reported in the literature, we cross-matched each of the ten VIPERS clusters with previously published catalogs overlapping the VIPERS W1 and W4 regions. A counterpart was considered a match when its projected separation from the VIPERS cluster center was $\leq 5$ arcmin and the redshift difference satisfied $\Delta z \leq 0.03$.  These thresholds account for the fact that different cluster catalogs often adopt different definitions of the cluster center (e.g., BCG position, luminosity peak, or density centroid), which can introduce offsets of a few arcminutes. They also accommodate the typical level of scatter between photometric and spectroscopic redshift estimates.

Table~\ref{tab:crossmatch} lists all previously cataloged systems found within our matching tolerances, including their coordinates, redshifts, angular separations from the VIPERS positions, and references to the source catalogs. Each matched system is shown on a separate row; VIPERS clusters that are associated with more than one cataloged system therefore appear on multiple rows. All ten VIPERS clusters have at least one previously reported cluster candidate within our adopted matching thresholds. Many of these matched clusters exhibit excellent agreement in both redshift and sky position (e.g., VIPERS01, VIPERS06, VIPERS09), while others show larger angular separations but comparable redshift estimates (e.g., VIPERS03). The presence of matched systems from multiple catalogs simply reflects the variety of survey depths and cluster-identification methods used in previous studies. Our spectroscopic selection therefore serves as an independent means of identifying these structures.

\section{Velocity Dispersion-Cluster Mass Relation}
\label{sec:sigmamass}

In this section, we investigate the scaling relation between the velocity dispersion and cluster mass ($\sigma$MR) for the VIPERS cluster sample. We first describe the methodology adopted for fitting the relation and then apply it to our data to derive the best-fit parameters: normalization ($\alpha$), slope ($\beta$), and intrinsic scatter ($\sigma_\mathrm{int}$).

\subsection{Methodology for Fitting the Scaling Relation}
\label{sec:method}

The probability distribution of a dependent variable $Y$ at fixed independent variable $X$ is assumed to follow a lognormal distribution (e.g., \citealp{Saro15,Simet17,Chiu20}):
\begin{equation} \label{eq:prob}
P(\log{Y}|X)= \frac{1}{\sqrt{2\pi\sigma^2_{\log{Y},X}}} 
\exp{\left[-\frac{\left(\log{Y} - \left<\log{Y}|X\right>\right)^2}{2\sigma^2_{\log{Y},X}}\right]},
\end{equation}
where the conditional mean relation is expressed as
\begin{equation} \label{eq:rich}
\left<\log{Y}|X\right> = \alpha + \beta \log{X}.
\end{equation}

The total variance in $Y$ ($\sigma^2_{\log{Y},X}$) at fixed $X$ includes contributions from measurement uncertainties in both $X$ and $Y$, as well as the intrinsic scatter in $Y$:
\begin{equation} \label{eq:var}
\sigma^2_{\log{Y},X} = \beta^2\sigma^2_{\log{X}} + \sigma^2_{\log{Y}} + \sigma^2_{\mathrm{int}},
\end{equation}
where $\alpha$ and $\beta$ denote the normalization and slope of the relation, respectively.  
The intrinsic scatter, $\sigma_{\mathrm{int}}$, quantifies the physical dispersion in $Y$ at fixed $X$ that cannot be explained by measurement uncertainties. In our analysis, $\sigma_{\mathrm{int}}$ is treated as a free parameter and is simultaneously fitted with $\alpha$ and $\beta$.

We estimate the parameters $\alpha$, $\beta$, and $\sigma_\mathrm{int}$ using the affine-invariant Markov Chain Monte Carlo (MCMC) sampler of \citet{Goodman10}, as implemented in the MATLAB package \texttt{GWMCMC}\footnote{\url{https://github.com/grinsted/gwmcmc}}, inspired by the Python package \texttt{emcee} \citep{Foreman13}.

\begin{figure*}\hspace{-0cm}    \includegraphics[width=1\linewidth]{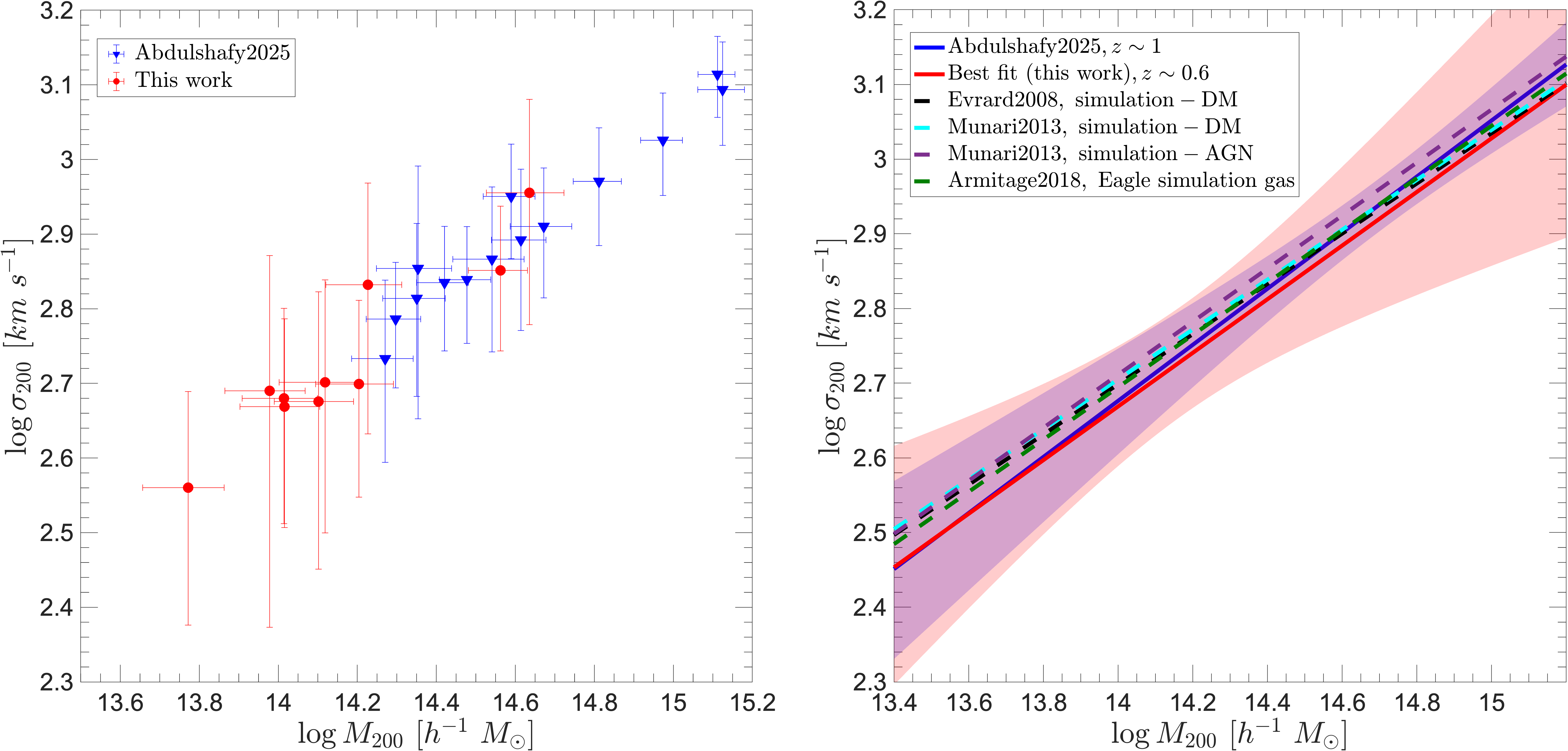} \vspace{-.5cm}
    \caption{Velocity dispersion $\sigma_{200}$ vs. virial mass $\mathrm{M}_{200}$. Left: the distribution of the cluster sample in the $\sigma_{200}$ - $\mathrm{M}_{200}$ plane, where red and blue points represent the results from our work and \citet{Abdulshafy25}, respectively. Right: the red and blue solid lines represent the best fit at $z=1$ for our work and \citet{Abdulshafy25}, respectively. For comparison, we include relations from $N$-body dark matter simulations, specifically those derived by \citet{Evrard08} (dashed black) and \citet{Munari13} (dashed cyan), as well as from simulations incorporating baryonic effects, including \citet{Munari13}-AGN (dashed violet) and \citet{Armitage18}-Eagle simulation (dashed green). The two shaded areas indicate $1\sigma$ uncertainties for our best-fit line and \citet{Abdulshafy25}}.
   \label{fig:SigM}
\end{figure*}

To fit the $\sigma$MR, we set $Y \equiv \sigma_{200}$ and $X \equiv M_{200}$. The best-fit parameters for the VIPERS sample, evaluated at a pivot mass of $M_\mathrm{piv} = 1.5\times10^{14}~h^{-1}~M_\odot$, are listed in Table~\ref{tab:fit}.  Figure~\ref{fig:SigM} shows the velocity dispersion $\sigma_{200}$ versus virial mass $M_{200}$. For comparison, we plot the $\sigma$MR obtained from \citet{Abdulshafy25} for a sample of 14 galaxy clusters take from the GOGREEN and GCLASS Surveys at $z\sim1$. 
The left panel displays the distribution of the cluster samples, where red points correspond to our VIPERS results and blue triangles to those from \citet{Abdulshafy25}. The right panel shows the best-fit $\sigma$MRs, with solid red and blue lines representing our fit and that of \citet{Abdulshafy25}, respectively. The shaded regions indicate the $1\sigma$ uncertainties on the best-fit relations for this work (red) and \citet{Abdulshafy25} (blue). For reference, we include theoretical predictions from dark matter–only $N$-body simulations by \citet{Evrard08} (dashed black) and \citet{Munari13} (dashed cyan), as well as relations from hydrodynamical simulations incorporating baryonic effects, including the AGN-feedback model of \citet{Munari13} (dashed violet) and the \textsc{Eagle} simulation of \citet{Armitage18} (dashed green). 

As shown in the left panel of Figure~\ref{fig:SigM}, the uncertainties in both $\sigma_{200}$ and $M_{200}$ are generally larger than those reported by \citet{Abdulshafy25}. This difference arises primarily from the smaller number of spectroscopically confirmed members per cluster in the VIPERS sample compared to the higher-multiplicity systems analyzed by \citet{Abdulshafy25}. With fewer galaxy velocities contributing to the dynamical estimates, the statistical uncertainty in $\sigma_{200}$ increases, which in turn propagates into the virial mass calculation since $M_{200}$ depends on $\sigma_{200}^2$ and $R_{200}$. The broader error bars in both quantities therefore reflect the combined effects of limited sampling and measurement propagation. As a result, the larger uncertainties in our dataset yield a correspondingly wider $1\sigma$ confidence region (shaded area) around the best-fit $\sigma$--$M_{200}$ relation.

\begin{table*}                               
\centering                                   
\begin{threeparttable}                                        
\caption{Fitting parameters for the velocity dispersion-cluster mass relation. }            
\label{tab:fit}         
\begin{tabular}{cccc} 
\toprule                                        
\multicolumn{4}{c}{Velocity Dispersion-Mass Relation ($\sigma$MR) at $\mathrm{M}_{\mathrm{piv}}=1.5\times10^{14}$ \hm} \\  
\midrule                                         &Normalization ($\alpha$) [\ks] &  Slope ($\beta$)  &   Intrinsic Uncertainty ($\sigma_{\mathrm{int}}$) \\                  
\midrule                                                                           
This Work  & 2.732 $\pm$ 0.063   & 0.359 $\pm$ 0.183    & 0.040 $\pm$ 0.072 \\
\citet{Abdulshafy25}  & 2.745 $\pm$ 0.028   & 0.373 $\pm$ 0.089    & 0.023 $\pm$ 0.024 \\
\citet{Evrard08}& $2.757\pm0.002$ & $0.336\pm0.003$ & $0.043\pm0.015$\\
\citet{Munari13}-DM& $2.764\pm0.002$ & $0.334\pm0.001$ & ---\\
\citet{Munari13}-CSF& $2.774\pm0.003$ & $0.355\pm0.003$ & ---\\
\citet{Armitage18}& $2.756\pm0.005$ & $0.350\pm0.010$ & ---\\
\bottomrule         
\end{tabular}              
\end{threeparttable}
\end{table*}

The right panel of Figure~\ref{fig:SigM} and Table~\ref{tab:fit} show that our best-fit relation is slightly offset toward lower $\sigma_{200}$ values at the high-mass end compared to \citet{Abdulshafy25}, while remaining consistent at lower masses. When compared to simulation-based predictions, our relation is similarly shifted toward lower $\sigma_{200}$ across the full mass range. The derived slope ($\beta = 0.359 \pm 0.183$) is in excellent agreement with the values predicted by cosmological simulations, particularly with the hydrodynamical results of \citet{Munari13} that include baryonic (CSF) physics. In contrast, the slope from \citet{Abdulshafy25}$\,$($\beta = 0.373 \pm 0.089$) is slightly steeper, lying marginally above the simulation results. These findings indicate that, despite the smaller and less populous cluster sample in VIPERS, our analysis recovers the same underlying $\sigma$--$M_{200}$ scaling, with a slope that is statistically consistent with theoretical expectations. The lower normalization of our fit likely reflects both the higher redshift of the VIPERS clusters and the larger measurement uncertainties associated with their limited spectroscopic membership.

Despite the relatively small number of spectroscopic members per cluster in our VIPERS sample, our results demonstrate that reliable dynamical measurements can still be obtained when using the FoG–GalWeight methodology in combination with the corrected virial mass estimator. As shown in Figure~\ref{fig:SigM}, the higher-mass clusters from \citet{Abdulshafy25}, which were derived from systems with substantially larger galaxy counts, smoothly extend the trend defined by our lower-mass VIPERS clusters, forming a complementary sequence across the full mass range. This continuity indicates that the virial mass estimator remains stable even for systems with limited spectroscopic sampling, provided that member identification is accurate. Moreover, the close agreement between our best-fit $\sigma$--$M_{200}$ relation and both simulation-based predictions and the \citet{Abdulshafy25} results confirms the robustness of our procedure. These findings reinforce that the FoG–GalWeight approach, coupled with the virial mass estimator, can reliably recover cluster dynamical properties even in the low-$N$ regime characteristic of high-redshift surveys.

\section{Conclusion} \label{sec:conc}

In this study, we analyzed galaxy clusters identified in the VIPERS spectroscopic survey within the redshift range $0.5 < z < 1.2$ using the FoG–GalWeight methodology. Cluster locations were first determined by detecting overdensities in redshift space associated with the Finger-of-God (FoG) effect. The centers of these clusters were then refined using the two-dimensional Adaptive Kernel Method (2DAKM), ensuring accurate determinations of their coordinates and redshifts.

Cluster membership was assigned using the GalWeight technique, which identifies galaxies in projected phase-space while minimizing contamination from interlopers. 
The resulting ten VIPERS clusters and their physical properties are summarized in Table~\ref{tab:pos}.
As shown in Figure~\ref{fig:PS}, the caustic profiles derived by GalWeight clearly trace the cluster boundaries and reproduce the expected “trumpet-shaped” structure characteristic of virialized systems in redshift space. The method successfully distinguishes true members even in systems with small spectroscopic samples, confirming its robustness at intermediate to high redshifts.

Using the identified members, we computed each cluster’s velocity dispersion ($\sigma_{200}$), virial radius ($R_{200}$), and virial mass ($M_{200}$) based on the virial mass estimator. The resulting cluster sample spans a mass range of $0.6\times10^{14}$–$4.3\times10^{14}~h^{-1}~M_\odot$ and a velocity dispersion range of $360$–$900$~km~s$^{-1}$, representing a mix of rich groups and low- to intermediate-mass clusters. To assess whether the VIPERS systems had been previously reported, we compared their positions and redshifts with those of clusters listed in published catalogs in the VIPERS W1 and W4 fields. All ten VIPERS clusters have at least one published cluster candidate falling within our positional and redshift tolerances, with several systems showing close agreement in both coordinates and redshift. This broad overlap with earlier detections provides external support for the FoG--GalWeight identifications while highlighting the variety of survey depths and cluster-finding techniques across catalogs.

The best-fit $\sigma$MR, derived using an MCMC-based regression analysis, returns $\log{\sigma_{200}} = (2.73 \pm 0.06) + (0.36 \pm 0.18)\,\log{M_{200}}$, with an intrinsic scatter of \(\sigma_\mathrm{int} = 0.04 \pm 0.07\). It is in excellent agreement with theoretical expectations from $N$-body and hydrodynamical simulations (e.g., \citealp{Evrard08,Munari13,Armitage18}). As shown in Figure~\ref{fig:SigM}, our results are consistent with the virial scaling predicted by simulations, confirming that velocity dispersion remains a reliable proxy for cluster mass even at $z \sim 1$. In comparison with the results of \citet{Abdulshafy25}, both datasets yield similar slopes, with our VIPERS clusters exhibiting a slightly higher normalization.

Overall, our analysis demonstrates that the FoG–GalWeight methodology provides a powerful framework for identifying and characterizing galaxy clusters in large spectroscopic surveys. The method yields robust membership assignments and dynamical mass estimates that agree closely with theoretical models. Importantly, our findings reaffirm that the velocity dispersion can be used directly to estimate cluster mass without the need for complex dynamical mass modeling. Future applications of this technique to larger spectroscopic samples, such as those from DESI and Euclid, will enable more precise measurements of cluster scaling relations and their evolution with redshift.

\section*{Acknowledgements}
GW gratefully acknowledges support from the National Science Foundation through grant AST-2347348.

\setcounter{section}{0}
\renewcommand{\thesection}{Appendix~\Alph{section}}
\renewcommand{\thefigure}{A}

\bibliography{Ref}{}
\bibliographystyle{aasjournal}

\end{document}